\documentclass[aps,pra,twocolumn,showpacs,superscriptaddress,longbibliography]{revtex4-1}
\usepackage{graphicx}

\usepackage{amsmath}

\usepackage{amssymb}

\usepackage{epstopdf}

\usepackage{amsfonts}

\usepackage{dcolumn}

\usepackage{multirow}

\usepackage{multirow}

\usepackage{CJK}
\usepackage{xcolor}

\usepackage[utf8]{inputenc}
\usepackage{esint}
\usepackage{float}

\makeatletter
\@ifundefined{textcolor}{}
{
 \definecolor{BLACK}{gray}{0}
 \definecolor{WHITE}{gray}{1}
 \definecolor{RED}{rgb}{1,0,0}
 \definecolor{GREEN}{rgb}{0,1,0}
 \definecolor{BLUE}{rgb}{0,0,1}
 \definecolor{CYAN}{cmyk}{1,0,0,0}
 \definecolor{MAGENTA}{cmyk}{0,1,0,0}
 \definecolor{YELLOW}{cmyk}{0,0,1,0}
}

\usepackage{subfigure}
\usepackage{epsfig}\usepackage{amsfonts}
\usepackage{mathrsfs}
\usepackage[toc,page,title,titletoc,header]{appendix}
\usepackage{amsmath}

\makeatother
\begin{document}
\title{Mean-field
spin-oscillation dynamics
 beyond the single-mode approximation
for a harmonically trapped spin-1 Bose-Einstein condensate}
\author{Jianwen Jie}
\address{Homer L. Dodge Department of Physics and Astronomy,
  The University of Oklahoma,
  440 W. Brooks Street,
  Norman,
Oklahoma 73019, USA}
\address{Center for Quantum Research and Technology,
  The University of Oklahoma,
  440 W. Brooks Street,
  Norman,
Oklahoma 73019, USA}
\author{Q. Guan}
\address{Homer L. Dodge Department of Physics and Astronomy,
  The University of Oklahoma,
  440 W. Brooks Street,
  Norman,
Oklahoma 73019, USA}
\address{Center for Quantum Research and Technology,
  The University of Oklahoma,
  440 W. Brooks Street,
  Norman,
Oklahoma 73019, USA}
\author{S. Zhong}
\address{Homer L. Dodge Department of Physics and Astronomy,
  The University of Oklahoma,
  440 W. Brooks Street,
  Norman,
Oklahoma 73019, USA}
\address{Center for Quantum Research and Technology,
  The University of Oklahoma,
  440 W. Brooks Street,
  Norman,
Oklahoma 73019, USA}
\author{A. Schwettmann}
\address{Homer L. Dodge Department of Physics and Astronomy,
  The University of Oklahoma,
  440 W. Brooks Street,
  Norman,
Oklahoma 73019, USA}
\address{Center for Quantum Research and Technology,
  The University of Oklahoma,
  440 W. Brooks Street,
  Norman,
Oklahoma 73019, USA}
\author{D. Blume}
\address{Homer L. Dodge Department of Physics and Astronomy,
  The University of Oklahoma,
  440 W. Brooks Street,
  Norman,
Oklahoma 73019, USA}
\address{Center for Quantum Research and Technology,
  The University of Oklahoma,
  440 W. Brooks Street,
  Norman,
Oklahoma 73019, USA}
\date{\today}

\begin{abstract}
Compared to single-component Bose-Einstein condensates,
spinor Bose-Einstein condensates display much richer dynamics.
In addition to density oscillations, spinor Bose-Einstein 
condensates exhibit
intriguing spin dynamics that is associated
with population transfer between
different hyperfine components. This
work analyzes the validity of the widely employed single-mode
approximation when describing the spin dynamics in
response to a quench of the system Hamiltonian.
The single-mode approximation
assumes that the different
hyperfine states all share the same time-independent
spatial mode.
This implies that
the resulting spin Hamiltonian only depends on the 
spin interaction strength and not on the density interaction strength.
Taking the spinor sodium Bose-Einstein
condensate in the $f=1$ hyperfine manifold 
as an example
and working within the mean-field theory framework,
it is found
numerically
 that the single-mode
approximation misses, in some parameter regimes, intricate details
of the spin and spatial dynamics. 
We develop a physical picture that explains the observed phenomenon.
Moreover, using that the population oscillations described by the
single-mode approximation enter into the effective potential felt by the 
mean-field spinor, we derive a semi-quantitative
condition for when dynamical mean-field induced
corrections to the single-mode approximation are relevant.
Our 
mean-field
results have implications for a variety of published and planned experimental
studies.
\end{abstract}
\maketitle

\section{Introduction}
\label{sec_introduction}
Spinor Bose-Einstein condensates (BECs) display rich physics including
spin domain formation, spin textures, 
topological excitations, and
non-equilibrium quantum dynamics~\cite{PR2012_Spinor_BEC:Ueda,RMP2013_Spinor_BEC:Ueda}.
Spin-1 BECs are most commonly realized using 
sodium or rubidium atoms in the $f=1$ hyperfine manifold.
Due to angular momentum conservation, the scattering of two
$m=0$ atoms into the $m= \pm 1$
hyperfine states provides a path toward entanglement 
generation~\cite{PRA1993Kitagawa,PRL1998Bigelow,PRLDuan2000,PRL2000Pu,riedel2010atom,gross2010nonlinear,strobel2014fisher,lucke2011twin,gross2011atomic,PRL2011Bookjans,ScienceYouLi2017,PANS2018Zou};
here $m$ denotes the projection
quantum number associated with the total angular momentum quantum number $f$
of a single atom.
This route for entanglement generation, which can be viewed as an analogue of
the four-wave mixing process in quantum optics, is behind a variety of
proposals aimed at spin squeezing and metrological 
gain~\cite{PRL2007Vengalattore,ma2011quantum,hamley2012spin,PRLSmerzi2015,PRL2016Linnemann,PRL2017Szigeti,RMP2018:Smerzi,PRAJie2019,PRA2019Qimin}.

In $^{23}$Na and $^{87}$Rb spin-1 BECs
($f=1$ manifold), the 
scattering length combination associated with the
spin interactions
is significantly smaller, in magnitude,
than that associated with 
the density interactions;
the ratio is approximately $28.1$~\cite{PRA2011Tiemann} and
$215$~\cite{RMP2013_Spinor_BEC:Ueda} for sodium and rubidium, respectively.
Correspondingly, the energy (time) scale
for the spin interactions
is smaller (larger) than for the 
density interactions.
This observation is the key behind the single-mode
approximation (SMA)~\cite{PRL1998Bigelow,PRA1999Pu,PRA2002_SMA:Yi,NJPZhang_2003,PR2012_Spinor_BEC:Ueda,RMP2013_Spinor_BEC:Ueda}, which---in the context
of a time-dependent situation---amounts to assuming that the 
shape, but not the amplitude, of the spatial density profile is frozen during the dynamics.
As a consequence, 
the spatial modes enter 
into the spin Hamiltonian only in the form
of the mean total density: a larger mean total density
corresponds to a larger, in magnitude, spin-dependent interaction
energy.

The SMA
has been employed at the quantum 
level~\cite{PRL1998Bigelow,PRA1999Pu,PRAJie2019,ScienceYouLi2017,NP2012_Amplification:Hamley,PRA2019Duan,NC2012Gerving} as well as at
the mean-field 
level~\cite{PRA2005Zhang,PRA2004Romano,PRA2005Sengstock,Kronj_ger_2008,gerbier2019}.
In the former, the spin Hamiltonian is treated fully
quantum mechanically.
In the latter, the mean value of the spin
components is considered,
resulting in a set of differential equations in terms of the
fractional population of the $m=0$ mode and the relative
phase that can be solved analytically.
Intriguingly, the set of differential equations can be reproduced
by defining a classical Hamiltonian in which
the relative phase and 
fractional  population
play the role
of the generalized coordinate and generalized momentum,
respectively~\cite{PRA2005Zhang}. This mapping allows one to visualize
the dynamics using phase portraits in two-dimensional
phase space.

The
 number of studies dedicated
to assessing the validity of the 
mean-field 
SMA quantitatively in 
experimentally realistic {\em{dynamical}} settings is rather 
small~\cite{PRL2007Lett,PRL2009Lett,PRA2014_MW_dressing:Zhao}. This paper
adds to this list and develops a simple framework for the emergence
 of dynamics beyond the SMA.
 Our work is related to Ref.~\cite{PRL2010Klempt}, which observed
 quantum fluctuation-driven resonances experimentally and
 analyzed these resonances  using the undepleted pump approximation, 
 which assumes that the predominantly occupied spin component
 remains macroscopically occupied during the spin dynamics.
 Just as Ref.~\cite{PRL2010Klempt}, we discuss a resonance effect.
The characteristics that distinguish the resonances discussed in the present work from
those discussed in Ref.~\cite{PRL2010Klempt} are:
(i) In our work, 
the time-dependent SMA solutions create an effective {\em{time-dependent}}
 potential (driving term) for each $m$ channel.
 The effective potentials seen by the $m = \pm 1$ components in Ref.~\cite{PRL2010Klempt},
 in contrast,
are {\em{time-independent}}.  If the effective potential felt by one
 component supports an excited eigenstate 
 whose energy is in resonance with the ground state energy of another channel,
 then 
coupling to an excited spatial mode becomes non-negligible.
Physically, the above energy condition corresponds to
a resonant scattering process in which
two $m=0$ atoms get scattered into an $m=+1$ and 
an $m=-1$ atom.
(ii)
The coupling between the
 spin and spatial degrees discussed in this paper 
 is  {\em{mean-field driven}} and not, as in Ref.~\cite{PRL2010Klempt},  {\em{quantum fluctuation driven}}.
 
 Our solutions to
 the coupled mean-field Gross-Pitaevskii equations show
 that
 the coupling between the spin and spatial degrees of freedom 
 develops {\em{dynamically}}, despite the fact that the initial state
 is well described within the mean-field SMA. 
   The resonance condition, which depends on the interactions, can be avoided by
   tuning the single-particle detuning between the $m=0$ and $m= \pm1$ atoms.
   Since the energy is conserved after the quench,
   the quench-induced dynamics discussed in our work is not accompanied by
   a
   relaxation to the ground state or the formation of
   (quasi-)static spin domains.
   Instead, spin structure develops and disappears as time proceeds.
  Our
  results
are expected to be useful for the interpretation of 
past, ongoing, and future experiments.

The remainder of this article is structured as follows.
Section~\ref{sec_theory} reviews the theoretical
framework employed.
Section~\ref{sec_results} presents results for a $^{23}$Na BEC
under external axially symmetric
harmonic confinement with fixed aspect ratio
for various single-particle energy shifts $q$ and
particle numbers $N$,
using an initial state
with vanishing magnetization ${\cal{M}}$. Beyond SMA physics is observed at the mean-field level.
Section~\ref{sec_experiment} explains the observed beyond SMA effects.
Finally, Sec.~\ref{sec_conclusions} summarizes our main results
and provides an outlook.

\section{Theoretical framework}
\label{sec_theory}

We consider a spin-1 BEC
consisting of $N$ mass $M$ atoms in an external harmonic
trap with angular frequencies 
$\omega_x$, $\omega_y$, and $\omega_z$.
In addition to the harmonic confinement, the BEC atoms 
are exposed to external magnetic
and microwave fields. The parameter $q$
in our equations below quantifies the strength of the
energy shift that arises
from the magnetic field induced  quadratic Zeeman shift 
and the
microwave field induced AC-Stark 
shift~\cite{PRA2006_MW_dressing:Gerbier,PRA2014_MW_dressing:Zhao}.
The linear Zeeman shift energy does not appear explicitly in the
equations since it can be eliminated by going to a rotating 
frame~\cite{NJPZhang_2003,RMP2013_Spinor_BEC:Ueda}.
Two different 
mean-field
descriptions are considered:
\begin{itemize}
\item {\em{Approach~A}}: A (2+2)-parameter 
mean-field SMA framework. This approach
amounts to solving two sets of equations
(one for the spatial and one for the spin degrees of freedom), both of which
depend on two parameters.
\item 
{\em{Approach~B}}: 
A 5-parameter coupled
Gross-Pitaevskii equations framework. This mean-field approach accounts for the
coupling of the spatial and spin degrees of freedom.
\end{itemize}
In both approaches,
the BEC is described by a three-component spinor
$\vec{\Psi}(\vec{r},t)=(\Psi_{+1}(\vec{r},t),
\Psi_{0}(\vec{r},t) ,
\Psi_{-1}(\vec{r},t))^T$;
however, the equations that govern $\Psi_{m}(\vec{r},t)$ 
differ for the two cases.

\subsection{\em{Approach A}}
The
mean-field SMA~\cite{PRL1998Bigelow,PRA1999Pu,PRA2002_SMA:Yi,NJPZhang_2003,PR2012_Spinor_BEC:Ueda,RMP2013_Spinor_BEC:Ueda}
assumes that
the $m=+1$, $0$, and $-1$ components
share the same spatial wave function 
$\psi_{\text{SMA}}(\vec{r}) \exp ( - \imath \epsilon t / \hbar)$,
where the Gross-Pitaevskii orbital $\psi_{\text{SMA}}(\vec{r})$ and the chemical
potential $\epsilon$ are solutions to the
stationary single-component 
Gross-Pitaevskii equation
\begin{eqnarray}
\bigg [
H_0
+ g_n (N-1) | \psi_{\text{SMA}}(\vec{r}) |^2 \bigg ]
\psi_{\text{SMA}}(\vec{r})
= \epsilon \psi_{\text{SMA}}(\vec{r})
\end{eqnarray}
with
\begin{eqnarray}
H_0 = \frac{- \hbar^2}{2M} \nabla_{\vec{r}}^2 + 
V_{\text{trap}}(\vec{r})
\end{eqnarray}
and
\begin{eqnarray}
V_{\text{trap}}(\vec{r})=
\frac{1}{2} M \left(
\omega_x ^2 x^2 +
\omega_y ^2 y^2 +
\omega_z ^2 z^2 \right) .
\end{eqnarray}
The Gross-Pitaevskii orbital $\psi_{\text{SMA}}(\vec{r})$ is assumed to
be normalized to one
and the density interaction
strength $g_n$
is defined through~\cite{PR2012_Spinor_BEC:Ueda,RMP2013_Spinor_BEC:Ueda} 
\begin{eqnarray}
g_n = \frac{4 \pi \hbar^2}{M} \frac{a_0 +2 a_2}{3},
\end{eqnarray}
where $a_0$ and $a_2$ are the $s$-wave scattering lengths 
for two colliding atoms with total spin angular momentum $F=0$
and $F=2$, respectively.
Assuming axially symmetric harmonic
confinement with $\omega_{\rho}=\omega_x=\omega_y$, $\psi_{\text{SMA}}(\vec{r})$
is governed by two dimensionless parameters, namely the
dimensionless mean-field strength 
$g_n(N-1)/(a_{\text{ho},z}^3 \hbar \omega_z)$ 
[$a_{\text{ho},z}=\sqrt{\hbar / (M \omega_z)}$]
and the trap aspect ratio $\lambda$,
$\lambda=\omega_z/\omega_{\rho}$.

The 
spinor components $\Psi_m(\vec{r},t)$ are then written as
\begin{eqnarray}
\label{eq_sma_ansatz}
\Psi_m(\vec{r},t)=\chi_{m}(t) \psi_{\text{SMA}}(\vec{r})
\exp ( - \imath \epsilon t / \hbar) ,
\end{eqnarray}
where the
$\chi_m(t)$, which govern the spin dynamics, are given by
$\chi_m(t)=\sqrt{\rho_{m}(t)} \exp \left[ \imath \theta_{m}(t) \right]$.
Here,
$\theta_m(t)$ and $\rho_m(t)$ 
denote the phase and fractional population,
normalized such that
$\rho_{+1}(t) + \rho_0(t) + \rho_{-1}(t) =1$,
of the $m$-th component.
The SMA
is argued to be applicable when the spin healing length $\xi_s$,
$\xi_s=\hbar / \sqrt{2 M |{c}_s|}$,
is larger than the size of the BEC~\cite{RMP2013_Spinor_BEC:Ueda,footnote_shl}.
Here, $c_s$ 
is the
spin interaction energy,
$c_s = g_s \overline{n}_{\text{SMA}}$,
where the spin interaction 
strength $g_s$~\cite{PR2012_Spinor_BEC:Ueda,RMP2013_Spinor_BEC:Ueda} is given by
\begin{eqnarray}
g_s = \frac{4 \pi \hbar^2}{M}\frac{a_2 - a_0}{3}
\end{eqnarray}
and
the mean density $\overline{n}_{\text{SMA}}$ 
by
$\overline{n}_{\text{SMA}} = N \int |\psi_{\text{SMA}}(\vec{r})|^4 d \vec{r}$.
Looking ahead, we also define the density interaction energy $c_n$,
$c_n = g_n {\overline{n}}_{\text{SMA}}$.

The equations that govern the fractional populations 
and phases
can be conveniently
written in terms of the magnetization
${\cal{M}}$, which is conserved throughout
the time dynamics, and the relative phase $\theta(t)$
between the spinors,
${\cal{M}} = \rho_{+1}(t) - \rho_{-1}(t)$
and
$\theta(t) = 2 \theta_0(t) -  \theta_{+1}(t) - \theta_{-1}(t) $.  
With these definitions, the coupled equations of motion 
read~\cite{PRA2005Zhang}
\begin{eqnarray}
\label{eq_eom1}
\frac{\hbar}{2c_s}\frac{d \rho_0(t)}{ d t} 
= -
  \rho_0(t) \sqrt{ [1 - \rho_0(t) ]^2 - {\cal{M}}^2 }
\sin( \theta(t))
\end{eqnarray}
and
\begin{eqnarray}
\label{eq_eom2}
&&-\frac{\hbar}{2 c_s} \frac{d \theta (t) }{dt} 
= -\frac{q}{c_s} +
1 - 2 \rho_0(t) \nonumber \\
&&+ 
\frac{1 - 3 \rho_0(t)+ 2[\rho_0(t)]^2 - {\cal{M}}^2}
     {  \sqrt{ [1 -\rho_0(t) ]^2 - {\cal{M}}^2 }}
\cos (\theta(t)).
\end{eqnarray}
Equations~(\ref{eq_eom1}) and (\ref{eq_eom2})
show that the mean-field spin dynamics
within the SMA is
fully determined
by two parameters, namely the ``dimensionless energy''
$q/c_s$ and the ``dimensionless time''
$t / (\hbar/c_s)$.
Since the static spatial mode and the spin dynamics
are decoupled and each
is governed by two parameters 
(throughout,
we are considering axially symmetric harmonic confinement), we refer to the mean-field
SMA as a (2+2)-parameter framework.
It should be noted, however, that ${\overline{n}}_{\text{SMA}}$,
which is determined by the static spatial mode, enters 
via the quantity $c_s$ into
the equations that determine the spin dynamics. 

\begin{figure}
\includegraphics[width=.35\textwidth]{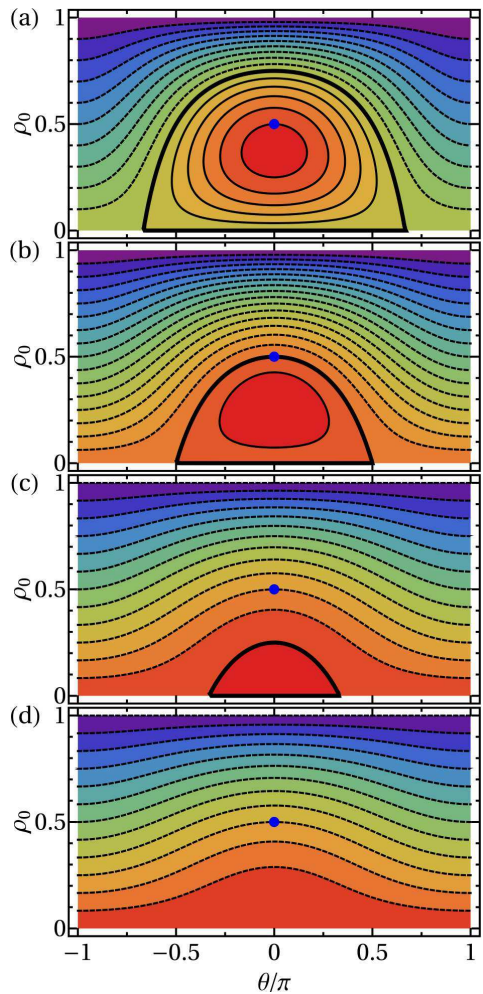}
\caption{Phase portraits illustrating the mean-field
SMA spin dynamics for ${\cal{M}}=0$ and
(a) $q/c_s=1/2$, (b) $q/c_s=1$,
(c) $q/c_s=3/2$, and (d) $q/c_s=2$.
The lines show equally spaced 
trajectories with fixed $E_s/c_s$;
the dimensionless energy spacing
is $1/20$, $1/16$, $1/8$, and $1/6$
in panels~(a)-(d), respectively.
The blue dots mark the initial conditions considered in
Sec.~\ref{sec_results};
they correspond to
$E_s/c_s=3/4$, $1$, $5/4$, and $3/2$
in panels~(a)-(d), respectively.
The separatrix that divides the periodic phase solutions
(solid lines) and running phase solutions
(dashed lines)
is shown by a
thick black line in panels~(a)-(c);
panel~(d) supports only
running phase and no periodic phase solutions.
}
\label{fig_phaseportrait}
\end{figure}

As already alluded to in Sec.~\ref{sec_introduction},
Eqs.~(\ref{eq_eom1}) and (\ref{eq_eom2})
can be interpreted as Hamilton's equations
of motion of a classical Hamiltonian, with $\theta(t)$
and $\rho_0(t)$ playing the roles of the generalized coordinate and 
associated generalized
momentum.
Within this framework, the spin energy $E_s$
is conserved~\cite{PRA2005Zhang}.
Figures~\ref{fig_phaseportrait}(a)-\ref{fig_phaseportrait}(d)
show the phase portrait 
for $q/c_s=1/2$, $1$, $3/2$, and $2$, respectively.
Given $\rho_0(0)$ and $\theta(0)$, the dynamics proceeds along a
fixed energy trajectory (lines
in Fig.~\ref{fig_phaseportrait}).
Depending on the initial conditions, the trajectories 
correspond to periodic phase 
solutions
(solid
lines in Fig.~\ref{fig_phaseportrait})
or running phase solutions
(dashed
lines in Fig.~\ref{fig_phaseportrait}).
For both classes of solutions,
$\rho_0(t)$ is characterized by a fixed
period and amplitude.
Our calculations in Sec.~\ref{sec_results} consider initial states with 
$\rho_0(0)=1/2$, $\rho_{+1}(0)=\rho_{-1}(0)=1/4$, and 
$\theta(0)=0$ (see blue dots in Fig.~\ref{fig_phaseportrait}).

\subsection{{\em{Approach B}}} 
Even if the initial state is
structureless and, e.g., 
well described by a Thomas-Fermi profile, spatial structure may develop
during the time dynamics. Indeed, the 
formation of spatial structure during spin
oscillation dynamics 
for spin-1
$^{23}$Na and $^{87}$Rb BECs
has been reported by several experimental 
groups~\cite{NP2005Chapman,PRL2010Klempt,PRA2013Klempt,PRL2013Raman}.
So far the theoretical modeling of this structure formation
has been, to the best of our knowledge, restricted to approximate
frameworks in the form of 
reduced dimensionality coupled mean-field Gross-Pitaevskii
simulations~\cite{PRA1999Pu}
or a stability analysis at the quantum level~\cite{PRL2010Klempt,klempt2010}.
The following paragraphs outline the time-dependent
coupled mean-field 
Gross-Pitaevskii 
equations framework, 
which allows for the 
coupling of the spin and spatial degrees of freedom and makes
no {\em{a priori}} assumption about the spatial dynamics of the  orbitals.

The time-dependent mean-field
Gross-Pitaevskii equations for the spinor
$\vec{\Psi}(\vec{r},t)$, which capture beyond single-mode physics,
can be conveniently written in
matrix form~\cite{PR2012_Spinor_BEC:Ueda,PRA2002_SMA:Yi,NJPZhang_2003,PRA1999Pu,PRA2000Pu}
\begin{eqnarray}
\label{eq_gp1}
\imath \hbar \frac{\partial \vec{\Psi}(\vec{r},t)}{\partial t}
= 
\left[ {\cal{L}} 
+ \underline{E}_{\text{shift}} 
+
\underline{V}_{\text{c}}(\vec{r},t)
\right]
\vec{\Psi}(\vec{r},t),
\end{eqnarray}
where
\begin{eqnarray}
\label{eq_gp2}
{\cal{L}}
= H_0 +
g_n (N-1) \sum_{m=0,\pm 1}| \Psi_m(\vec{r},t)|^2;
\end{eqnarray}
${\underline{E}}_{\text{shift}}$ is a diagonal matrix with diagonal
elements $q$, $0$, and $q$; and
\begin{widetext}
\begin{eqnarray}
\label{eq_gp2b}
\underline{V}_{\text{c}}(\vec{r},t)
= g_s(N-1) \times
\nonumber \\
\left( \begin{array}{ccc}
\sum_{m=\pm 1,0}|\Psi_{m}(\vec{r},t)|^2-2|\Psi_{-1}(\vec{r},t)|^2  & 
\Psi_0(\vec{r},t)\left[\Psi_{-1}(\vec{r},t)\right]^* & 
0 \\
\left[\Psi_0(\vec{r},t)\right]^* \Psi_{-1}(\vec{r},t) & 
\sum_{m=\pm 1,0} |\Psi_{m}(\vec{r},t)|^2 - |\Psi_{0}(\vec{r},t)|^2  & 
\left[\Psi_0(\vec{r},t)\right]^* \Psi_{+1}(\vec{r},t) \\
0 & 
\Psi_0(\vec{r},t) \left[\Psi_{+1}(\vec{r},t)\right]^* & 
\sum_{m= \pm 1,0}|\Psi_{m}(\vec{r},t)|^2 -2 |\Psi_{+1}(\vec{r},t)|^2 
\end{array}
\right)
.
\end{eqnarray}
\end{widetext}
If the spin interactions vanish (i.e., if 
$g_s=0$), 
then the solutions are independent of the
coupling matrix $\underline{V}_{\text{c}}(\vec{r},t)$
[see Eqs.~(\ref{eq_gp1}) and (\ref{eq_gp2b})]. 
The pattern formation discussed in Sec.~\ref{sec_results}
crucially depends on the kinetic energy contributions
in Eq.~(\ref{eq_gp1}) [see also
Eq.~(\ref{eq_gp2})], i.e., 
treatment of the coupled
mean-field Gross-Pitaevskii
equations within
the Thomas-Fermi approximation
yields qualitatively different results than treatment of the full
coupled mean-field equations.

As discussed earlier, the spatial dynamics and 
the spin dynamics
within the mean-field SMA depend 
each on two dimensionless parameters,
namely, 
$g_n(N-1)/(a_{\text{ho},z}^3 \hbar \omega_z)$ 
and $\lambda$ for the
spatial degrees and $q/c_s$
and $t/(\hbar/c_s)$ for the spin degrees.
The coupled Gross-Pitaevskii equations 
[see Eqs.~(\ref{eq_gp1})-(\ref{eq_gp2})], in contrast,
depend on five dimensionless parameters:
$g_n(N-1)/(a_{\text{ho},z}^3 \hbar \omega_z)$, $\lambda$,
$q/c_s$, $t/(\hbar/c_s)$,
and $g_n/g_s$. The ratio $g_n/g_s$ ``connects'' the spatial and
spin degrees of freedom.
The energy scales 
$\hbar \omega_{\rho}$, $\hbar \omega_z$, and 
$g_n(N-1)/a_{\text{ho},z}^3$ are---for typical experimental parameters---significantly
larger than the energy scale
$c_s$. This suggests that the dynamics that is resulting from these
energy scales
is faster than the ``low-energy'' spin population dynamics.
In fact, the mean-field SMA
assumes that the dynamics introduced by these high-energy
scales is so fast that it can be safely averaged out.
However, the coupling between the low- and high-energy
degrees of freedom can, 
at least in principle, lead to an energy transfer between the
associated  
degrees of freedom. While direct comparisons
between Gross-Pitaevskii simulation results and experimental
data were not made, Ref.~\cite{NP2005Chapman}
attributed the experimentally observed 
damping of the spin oscillations 
for $^{87}$Rb (positive $q$) to this energy transfer and, associated with it,
the breakdown of the mean-field SMA.

\section{Numerical Results}
\label{sec_results}
We consider a spin-1 $^{23}$Na condensate
with $a_0=48.91a_{B}$ and $a_2=54.54a_{B}$~\cite{PRA2011Tiemann},
where $a_B$ denotes the Bohr radius.
To prepare the initial state, we imagine the following
procedure.
First, all atoms are loaded into the $m=-1$ state
of the $f=1$ hyperfine manifold in the presence of a small
magnetic field.
Second, a radio-frequency pulse is applied to prepare a state
with population fractions of $1/4$ and $1/2$
in the $m= \pm 1$ and $m=0$
hyperfine states of the $f=1$ manifold~\cite{PRA2014_MW_dressing:Zhao,PRL2007Lett}.

\begin{figure}
\includegraphics[width=.38\textwidth]{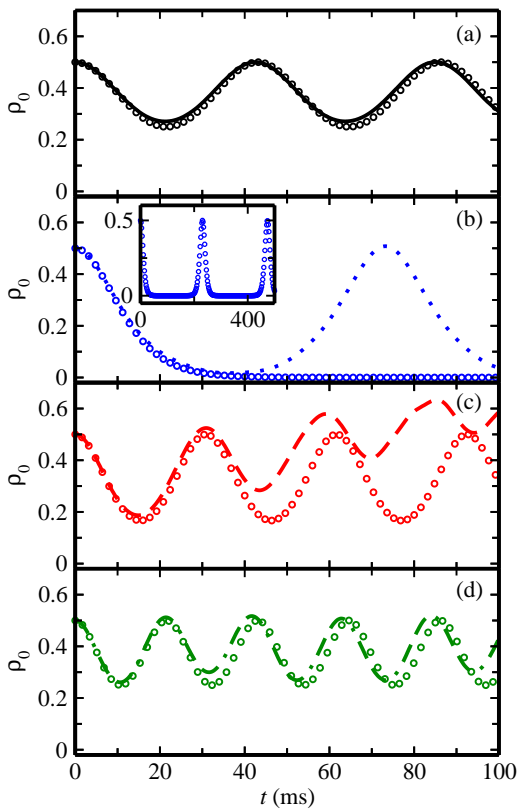}
\caption{Comparison of 
the time-dependent fractional population
$\rho_0(t)$ obtained using  the coupled mean-field Gross-Pitaevskii 
equations 
(lines; {\em{Approach B}})
and mean-field SMA (open circles; {\em{Approach A}}).
The coupled Gross-Pitaevskii equations simulations are performed for a sodium spin-1
BEC with 
$N=40000$ under external harmonic
confinement
characterized by  $\omega_{\rho}= 2 \pi \times 70$~Hz 
and $\omega_z=2 \pi \times 150$~Hz
and interaction energies
$c_s /h \approx 12.5$~Hz and $c_n / h \approx 350$~Hz.
The single-particle energy scale
$q$ is 
(a) $q/c_s=1/2$,
(b) $q/c_s=1$,
(c) $q/c_s=3/2$,
and
(d) $q/c_s=2$;
this corresponds to
$q/h\approx 6.23$~Hz, $12.5$~Hz, $18.7$~Hz, and $24.9$~Hz,
respectively.
The inset in panel~(b) shows the mean-field SMA result
for larger $t$.
}
\label{fig_spinoscillation}
\end{figure}

The lines in Fig.~\ref{fig_spinoscillation}
show the fractional population  $\rho_0(t)$ as a function
of time, obtained by solving 
the coupled Gross-Pitaevskii equations 
for $N=40000$ and an axially symmetric harmonic confinement with moderate
aspect ratio of $\lambda=15/7 \approx 2.143$ for four different $q$ values. 
For comparison, the mean-field SMA results are shown by open circles.
For the smallest and largest $q/c_s$ considered
[Figs.~\ref{fig_spinoscillation}(a) 
and \ref{fig_spinoscillation}(d) 
are for 
$q/c_s=1/2$ and $2$, respectively], the mean-field SMA describes
the full mean-field
spin oscillation dynamics fairly accurately. 

For $q/c_s=1$, in contrast, the results for the
coupled
Gross-Pitaevskii equations ({\em{Approach B}})
display spin oscillations
with notably 
smaller period than the results for 
the mean-field SMA ({\em{Approach A}})
[$73$~ms versus $231$~ms, see the inset of Fig.~\ref{fig_spinoscillation}(b)].
For this parameter combination,
the coupling between the spin and spatial degrees of freedom
speeds the spin oscillation dynamics up significantly, i.e., the
mean-field Gross-Pitaevskii equations
framework predicts a smaller period than the
mean-field SMA framework.
In the classical phase
portrait,
the initial
state for $q/c_s=1$ is
located on the separatrix.
As can be seen from Fig.~\ref{fig_phaseportrait}(b), this implies
that the fractional population $\rho_0(t)$
is equal to zero half-way through the first oscillation of the
fractional populations.
The coupled Gross-Pitaevskii equations, in contrast,
result in a small non-zero $\rho_0(t)$ half-way through the first oscillation of the
fractional populations; specifically,
the value of $\rho_0(t)$ for $t =36.4$~ms
is approximately $0.019$, corresponding to
about $700$ atoms in the
$m=0$ hyperfine state. For such a low-atom BEC component,
quantum fluctuations of the spin and/or spatial   
degrees of freedom may play a non-negligible role,
suggesting that the applicability of the coupled Gross-Pitaevskii
equations 
needs to be assessed carefully. 

For $q/c_s=3/2$ [see Fig.~\ref{fig_spinoscillation}(c)], the coupled Gross-Pitaevskii
equations and the mean-field SMA both display 
spin population oscillations of---roughly---the same period.
Intriguingly, however, the coupling between the spin and spatial degrees
of freedom that is accounted for by the coupled Gross-Pitaevskii equations leads to 
a damping as well as an overall upward drift 
of the spin oscillation amplitude during the first few cycles.
This upward drift is not captured by the mean-field SMA, 
which predicts fractional population oscillations with
constant amplitude and period.

\begin{widetext}

\begin{figure}
\includegraphics[width=1.0\textwidth]{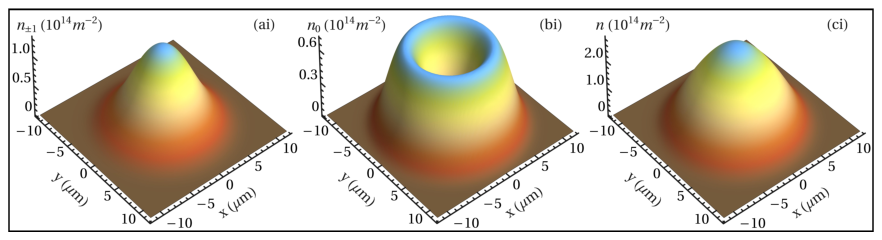}
\includegraphics[width=1.0\textwidth]{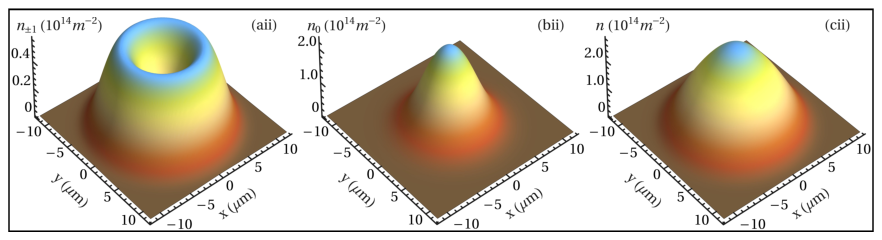}
\caption{Snapshots of the integrated spatial density for
a $^{23}$Na spin-1 BEC with $N=40000$ 
under external harmonic confinement with
$\omega_{\rho}= 2 \pi \times 70$~Hz and $\omega_z= 2 \pi \times 150$~Hz. 
The parameters are the same as those in
Fig.~\ref{fig_spinoscillation}(c), i.e., 
$q/c_s=3/2$, $c_s/h \approx 12.5$~Hz, and $c_n/h \approx 350$~Hz.
The top row [panels~(ai)-(ci)] 
and bottom row [panels~(aii)-(cii)]
are for $t=38$~ms and 
$50$~ms, respectively.
The first, second, and third columns
show the integrated densities
$n_{\pm 1}(x,y,t)=N \int_{-\infty}^{\infty} |\Psi_{\pm 1}(\vec{r},t)|^2 d z$,
$n_0(x,y,t) = N \int_{-\infty}^{\infty} |\Psi_{0}(\vec{r},t)|^2 d z$,
and
$n(x,y,t) = N \sum_{m=0,\pm 1}\int_{-\infty}^{\infty} |\Psi_{m}(\vec{r},t)|^2 d z$,
respectively.
}
\label{fig_density}
\end{figure}

\end{widetext}

Figure~\ref{fig_density}
shows selected integrated density profiles for the same parameters as
those used in Fig.~\ref{fig_spinoscillation}(c), i.e., for the case where the 
fractional population $\rho_0(t)$
drifts upward with time. Recall, this happens for $q/c_s=3/2$
and an initial state that is far away from the separatrix
[see Fig.~\ref{fig_phaseportrait}(c)].
The first two rows of Fig.~\ref{fig_density}
are for $t=38$~ms (after a bit more than 
one spin population oscillation) and the last two rows for
$t=50$~ms (after about one and a half spin population
oscillations).
For both times, we see that the total integrated densities 
[see Figs.~\ref{fig_density}(ci)
and
\ref{fig_density}(cii)] 
are close to what would be expected
within the SMA and
that the integrated densities for the
spinor components
[see Figs.~\ref{fig_density}(ai), 
\ref{fig_density}(bi), 
\ref{fig_density}(aii), and 
\ref{fig_density}(bii)]
display non-trivial structure, i.e., have much less resemblance with a 
Thomas-Fermi profile.
Figure~\ref{fig_density} shows that the 
structures
of $n_0(x,y,t)$ for $t=38$~ms and of $n_{\pm 1} (x,y,t)$ for $t=50$~ms
are quite similar. This observation is more general: We find that
the ``deformations'' of the subcomponent densities $n_0(\vec{r},t)$ and 
$n_{\pm 1}(\vec{r},t)$,
which combine to a Thomas-Fermi
like total density $n(\vec{r},t)$,
oscillate ``out of phase''.
This oscillatory structure in the subcomponent densities
can clearly not be described within the SMA.
We also solved the coupled mean-field Gross-Pitaevskii
equations within the Thomas-Fermi approximation, which neglects
the kinetic energy terms. This approximation
yields qualitatively different densities than those displayed
in Fig.~\ref{fig_density}. This shows that the observed 
structure formation depends sensitively on the interplay
between the various energy terms in the
coupled mean-field Gross-Pitaevskii equations.

\begin{figure}
\includegraphics[width=.35\textwidth]{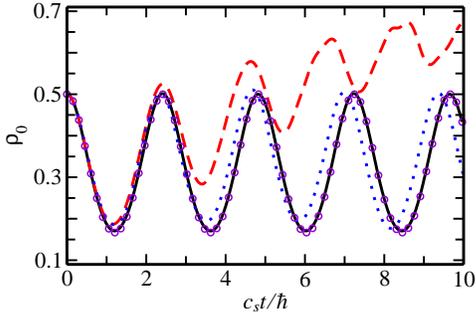}
\hspace*{1in}
\caption{Fractional population $\rho_0(t)$, determined
by the coupled Gross-Pitaevskii 
equations
for a sodium spin-1
BEC under external harmonic
confinement
characterized by  $\omega_{\rho}= 2 \pi \times 70$~Hz 
and $\omega_z= 2 \pi \times 150$~Hz.
The 
black solid, red dashed, and blue dotted
lines are for $N=10000$, $40000$, and $80000$, respectively.
The $q$ value is adjusted for each $N$ such that $q/c_s=3/2$ for all
three $N$ considered.
For comparison, the open circles show the result
from the calculations within the mean-field SMA; 
since the time is plotted in dimensionless units, the mean-field 
SMA result is independent of $N$.
The open circles are essentially indistinguishable 
from the coupled Gross-Pitaevskii equations results for $N=10000$.
}
\label{fig_validity_sma2}
\end{figure}

The break-down of the SMA
can be illustrated in a complementary approach, which 
relies on the fact that the
mean-field SMA framework is fully governed by
two dimensionless parameters, namely
$q/c_s$ and $t/ (\hbar / c_s)$.
The open circles in Fig.~\ref{fig_validity_sma2}
show the fractional population $\rho_0(t)$,
obtained within the mean-field SMA,
as  a function of the dimensionless time for 
$q/c_s = 3/2$ 
[same data as in Fig.~\ref{fig_spinoscillation}(c)].
For comparison, the red dashed, black solid, and blue 
dotted curves
show the coupled mean-field
Gross-Pitaevskii equations results for $N=40000$
[same data as in Fig.~\ref{fig_spinoscillation}(c)],
$N=10000$, and $N=80000$, respectively.
In all cases, the trap frequencies $\omega_{\rho}$ and $\omega_z$
and coupling strengths $g_s$ and $g_n$
are the same as before. However, the value of $q$ is
adjusted such that $q/c_s=3/2$ for all three $N$ values considered.
Figure~\ref{fig_validity_sma2} shows
that the $N=10000$ results are essentially on top of the 
SMA results, that the $N=80000$ oscillations have a slightly 
reduced oscillation period and amplitude, and that the $N=40000$
data display---as discussed in detail above---notable deviations
from the SMA result.
The ratio
$\xi_s/R_{\text{TF},z}$ is
$\approx 1.51$, $\approx 0.84$, and $\approx 0.63$
for $N=10000$, $40000$, and $80000$, respectively.
Thus, the reliability of the 
mean-field SMA
is not solely governed by the ratio between the spin healing length 
and the Thomas-Fermi radii.

Focusing on initial states with fractional populations
of
$1/4$, $1/2$, and $1/4$ for the 
$m=+1$, $m=0$, and $m=-1$ hyperfine levels and
vanishing relative phase,
Figs.~\ref{fig_spinoscillation}-\ref{fig_validity_sma2} identify two regimes
where the mean-field SMA ({\em{Approach~A}})
and coupled mean-field Gross-Pitaevskii
equations ({\em{Approach~B}}) yield different results.
(i) Unlike {\em{Approach~A}}, {\em{Approach~B}} reveals an overall drift
of the oscillating fractional population $\rho_0(t)$, which is accompanied
by time-dependent non-Thomas-Fermi like pattern formation.
(ii) The time periods of the spin oscillations 
predicted by {\em{Approach~A}} and {\em{Approach~B}}
deviate significantly
for $q/c_s \approx 1$, i.e.,
in the regime where the initial state is located on the 
separatrix.
The next section develops a theory understanding of regime~(i).

\section{Physical picture}
\label{sec_experiment}

The effective potential picture developed in this section 
based on the 
time-dependent Gross-Pitaevskii equations [Eqs.~(\ref{eq_gp1})-(\ref{eq_gp2b})] is conceptually
similar to the effective potential picture developed in
 Ref.~\cite{PRL2010Klempt}, using a framework that accounts for quantum fluctuations. 
 In that work, the $m = \pm 1$ modes are initially empty and the effective potentials are time independent.
 In our work, in contrast, the $m = \pm 1$ modes are initially macroscopically occupied and the 
 effective potentials are time dependent.

In the absence of interactions and positive $q$,
two colliding $m=0$ atoms must have
an ``extra" energy of $2q$ to
scatter into an $m=+1$ atom and an $m=-1$ atom. 
If $q$ is negative, then an $m=+1$ atom and an $m=-1$ atom
must have an ``extra" energy of $2q$ 
 to
scatter into two $m=0$ atoms.
To streamline the discussion, we assume in the remainder
of this section that $q$ is positive; the arguments
can be readily extended to the negative $q$ case.
Since the $m=0$ channel needs---in the absence of interactions---an ``extra" energy of $2q$
to be in resonance with the $m=\pm 1$ channels, we 
anticipate that---in the presence of interactions---the 
drifting occurs when there exists an 
excited state in the $m=0$ 
channel that is in resonance with the 
ground states of the $m = \pm 1$ channels. 
In this resonant regime,
population transfer to the excited state can occur,
leading to a density deformation (physics beyond the SMA).
Since the different $m$ channels are coupled in the presence of interactions,
the ``excited" and ``ground" states just referred to are associated with effective
potentials that neglect the coupling between channels. 

Our  semi-quantitative estimate starts with 
Eqs.~(\ref{eq_gp1})-(\ref{eq_gp2b}). 
We 
work within the SMA to evaluate
the effective potential $\underline{V}_{\text{eff}}(\vec{r},t)$,
\begin{eqnarray} 
\label{eq_veff}
\underline{V}_{\text{eff}}(\vec{r},t)=
V_{\text{trap}}(\vec{r}) \underline{I} +
\underline{E}_{\text{shift}}+ \underline{V}_{\text{nl}}(\vec{r},t),
\end{eqnarray}
where
\begin{eqnarray}
\underline{V}_{\text{nl}}(\vec{r},t)=\underline{V}_c(\vec{r},t)+g_n(N-1) \sum_m |\Psi_m(\vec{r},t)|^2 \underline{I}, 
\end{eqnarray}
and
$\underline{I}$ denotes the $3 \times 3$ identity matrix. 
Specializing to the case where 
$|\Psi_{+1}(\vec{r},t)|=|\Psi_{-1}(\vec{r},t)|$, 
we find
\begin{eqnarray}
\underline{V}_{\text{nl}}(\vec{r},t)=
(N-1) |\psi_{\text{SMA}}(\vec{r})|^2 \underline{V}_{\text{drive}}(t),
\end{eqnarray}
where
\begin{eqnarray}
\underline{V}_{\text{drive}}(t)= 
g_n \underline{I} + g_s \underline{V}_{\text{diag}}(t) + g_s \underline{V}_{\text{off-diag}}(t).
\end{eqnarray}
Here, 
$\underline{V}_{\text{diag}}(t) $ is a diagonal matrix
with elements $\rho_{0}(t)$, 
$1-\rho_{0}(t)$, and  $\rho_{0}(t)$;
\begin{eqnarray}
\underline{V}_{\text{off-diag}}(t)=
\left( \begin{array}{ccc}
0&  d_{-1}(t) & 0\\
d_{-1}^*(t)  & 0 & d^*_{+1}(t)  \\
0 &  d_{+1}(t)  &0
\end{array}
\right);
\end{eqnarray}
and
\begin{eqnarray}
d_{\pm1}(t) =  \sqrt{\frac{\rho_0(t)[1-\rho_0(t)]}{2}} 
e^{ \imath[\theta_0(t)-\theta_{\pm1}(t)  ]}.
\end{eqnarray}
To proceed, we treat the time $t$ as an adiabatic 
parameter, neglect 
$\underline{V}_{\text{off-diag}}(t)$, and solve the
linear Schr\"odinger equation for
the effective potential $\underline{V}_{\text{eff}}(\vec{r},t)$.
Explicitly, the $m=\pm1$ and  $m=0$ effective potentials
read as
\begin{eqnarray}
V_{\text{eff},\pm1}(\vec{r},t)= \nonumber \\
V_{\text{trap}}(\vec{r})+ q+  \left[ g_n+g_s \rho_0(t) \right](N-1) |\psi_{\text{SMA}}(\vec{r})|^2
\end{eqnarray}
and
\begin{eqnarray}
V_{\text{eff},0}(\vec{r},t)= \nonumber \\
V_{\text{trap}}(\vec{r})+   \left[ g_n+g_s (1-\rho_0(t)) \right](N-1) |\psi_{\text{SMA}}(\vec{r})|^2.
\end{eqnarray}
Since the effective potentials depend on time through $\rho_0(t)$, 
the resonance condition changes with time.
To estimate the resonance condition, we use the effective
potential curves for $t=0$,
i.e., we set $\rho_0(t)$ equal to $1/2$;
this implies that the coupling constants are equal to $g_n+g_s/2$ in all three channels.
While there is some arbitrariness in this choice,
the physical picture is not impacted by this.
When the 
$m=0$ channel
supports an excited
state 
that lies $2q$ above the ground state energy of the $m\pm1$ channels,
the resonance condition
is
fulfilled.
A more rigorous treatment would include the coupling terms and 
might consider a time average. 

For the parameters of Fig.~\ref{fig_spinoscillation}, our approximate formalism
yields that the drifting should occur at $q_{\text{res}} \approx 21 $~Hz.
This estimate agrees quite well with the result obtained by solving the coupled Gross-Pitaevskii 
equations, which shows that the drifting is maximal
for $q \approx 19$~Hz.
We also estimate the resonance conditions for 
$N=10^4$ and $N=8 \times 10^4$, using the same
parameters as in Fig.~\ref{fig_validity_sma2}. 
Our approximate formalism yields
$q_{\text{res}} \approx 31$~Hz and $q_{\text{res}} \approx 17$~Hz, 
respectively, in
good agreement with the observed maximal drifting for
 $q \approx 30$~Hz and  $q \approx 17$~Hz.
 To estimate  $q_{\text{res}}$, we used the Thomas-Fermi approximation~\cite{pethick_and_smith}. 
 This implies that the effective potential $\underline{V}_{\text{eff}}(\vec{r},t)$ is constant
 in the regime where the density $|\psi_{\text{SMA}}(\vec{r}) | ^2$, estimated
 within the Thomas-Fermi approximation using the coupling constant $g_n+g_s/2$, 
 is finite and equal to the
 harmonic
 oscillator potential otherwise. As a consequence, the energy of the first 
 ``radially" excited state
 (excitation predominantly located along the $\rho$-coordinate) 
sits by an energy that is comparable to the Thomas-Fermi energy above the ground state energy.
Importantly, our approximate framework also predicts higher-lying resonances,
corresponding to higher-lying excited states that are supported by the effective 
potentials $V_{\text{eff},m}(\vec{r},t)$, and higher-order resonances.
Our numerical solutions to the
coupled  Gross-Pitaevskii equations confirm these predictions. 
We checked the predictive power of our approximate framework for
about 10 different $N$, $\omega_{\rho}$, $\omega_z$, $g_n$, and $g_s$
parameter combinations
and found that it predicts the first drifting condition,
i.e., the value of $q_{\text{res}}$, at roughly the 15~\% level.
We emphasize that the drifting is not only observed for 
sodium spin-1 BECs but also for spin-1 BECs with larger $g_n/g_s$.
Moreover, analogous effects are anticipated to occur for higher-spin BECs.

\section{Conclusions}
\label{sec_conclusions}

This paper investigated the applicability of the SMA for a 
quenched spin-1 BEC. Specifically, the 
system Hamiltonian was quenched at time zero and the 
subsequent time evolution
was analyzed.
All figures presented show results for
a $^{23}$Na spin-1 BEC under axially symmetric harmonic
confinement with moderate aspect ratio and 
atom numbers of typical experiments.
This system has been used extensively
to study spin oscillations and published experimental 
data~\cite{PRA2014_MW_dressing:Zhao,PRL2007Lett,PRA2018Wrubel} 
have been 
interpreted
as 
validating the SMA.
Relying on the 
applicability of the SMA,
follow-up work used $^{23}$Na spin-1 BECs to study (quantum) phase transitions
that are supported by
the spin 
Hamiltonian~\cite{PRL2009Lett,PRA2017Vinit,PRL2011Bookjans,ScienceYouLi2017,PRA2019Duan,duan2020}.
At the same time,
some
experimental observations,
which cannot be readily reconciled with the validity of the
SMA,
have been reported~\cite{PRL2010Klempt,NP2005Chapman,PRA2019Duan,PRL2011Bookjans}, not
only for $^{23}$Na spin-1 BECs but also for $^{87}$Rb
spin-1 BECs. 
Quite generically, 
it is said that the SMA should become better as the 
ratio between the density- and spin-interaction strengths
increases. This ratio is $28.1$ for the $f=1$ manifold of $^{23}$Na~\cite{PRA2011Tiemann}
and
$215$ for the $f=1$ manifold of $^{87}$Rb~\cite{RMP2013_Spinor_BEC:Ueda}.
Spin-1 BECs can also be realized 
using the 
$f=2$ 
or larger $f$ hyperfine manifolds,
provided the $m=\pm 2, \cdots,\pm f$ atomic levels
are unoccupied.
The $f=2$ manifold of $^{87}$Rb has, e.g., been used
to realize a spin-1 system~\cite{PRL2016Linnemann,Widera_2006}.

Section~\ref{sec_results}
showed that the solutions to the 
time-dependent mean-field Gross-Pitaevskii equations 
yield, for certain parameter combinations, spin oscillations that 
deviate appreciably from those obtained within the mean-field SMA.
The drifting
of the spin oscillations were interpreted as a key signature
of beyond mean-field SMA physics.
Section~\ref{sec_experiment}
showed that the drifting 
occurs, assuming positive $q$, when the 
energy of the excited state supported by 
the effective 
$m=0$ mean-field potential curve is in resonance with the ground state supported by the $m = \pm 1$ effective
mean-field potential curves:
When the excited spatial mode has just the right energy,
 two excited $m=0$ atoms are in resonance with 
a pair of $m = \pm 1$ atoms, providing a coupling mechanism that leads
to spatial deformations that are not described by the mean-field SMA orbital.
An analogous argument applies to negative $q$.
The {\em{dynamical mean-field driven}} resonance effect discussed in this paper,
which exists for positive and negative $q$, complements earlier
work that experimentally measured and theoretically analyzed
 {\em{quantum-fluctuation driven}} resonances~\cite{PRL2010Klempt}.

Our predictions have a wide range of implications for, e.g.,
the calibration of effective Rabi coupling strengths and
proposals that are aimed at metrological 
gain~\cite{PRL2007Vengalattore,ma2011quantum,hamley2012spin,PRLSmerzi2015,PRL2016Linnemann,PRL2017Szigeti,RMP2018:Smerzi,PRAJie2019,PRA2019Qimin}.
If the spatial degrees of freedom cannot be treated as ``stiff'',
describing the quantum properties of the spin degrees of freedom will be 
significantly more involved.
To minimize the coupling between
the spatial and spin degrees of freedom,
in practice
one will likely want to work away from the regime where 
the resonances that were predicted in this work occur.
Taking an alternative viewpoint, the physics in the strongly-coupled
regime may 
be an interesting subject in itself.

\section{Acknowledgement}
\label{acknowledgement}
Support by the National Science Foundation through
grant numbers
PHY-1806259 (JJ, QG, DB) and 
PHY-1846965 (CAREER; SZ, AS)
is gratefully acknowledged.
This work used  the OU
Supercomputing Center for Education and Research
(OSCER) at the University of Oklahoma (OU).

\end{document}